\documentclass[aps,prb,groupedaddress,amsmath]{revtex4}

\usepackage{amssymb}
\usepackage{amsmath}
\usepackage{graphicx}
\usepackage{amsbsy}
\usepackage{color}

\newcommand{\bq}{\begin{eqnarray}}
\newcommand{\eq}{\end{eqnarray}}

\begin{document}
\title{Thermodynamic consistency of energy and virial routes:
 An exact proof within the linearized Debye--H\"uckel theory}

\author{Andr\'es Santos}
\email{andres@unex.es;
URL:http://www.unex.es/fisteor/andres/}
\affiliation{Departamento de F\'isica, Universidad de Extremadura,
E-06071 Badajoz, Spain}
\author{Riccardo Fantoni}
\email{rfantoni@ts.infn.it}
\affiliation{National Institute for Theoretical Physics (NITheP), Stellenbosch 7600, South Africa}
\author{Achille Giacometti}
\email{achille@unive.it}
\affiliation{Dipartimento di Chimica Fisica, Universit\`a di Venezia,
Calle Larga S. Marta DD2137, I-30123 Venezia, Italy}
\date{\today}

\begin{abstract}
The linearized Debye--H\"uckel theory for liquid state is shown to provide thermodynamically consistent virial and energy routes
for any  potential and for any dimensionality. The importance of this result for bounded potentials
is discussed.
\end{abstract}

\maketitle
Integral equations of liquid theory always involve some approximate closure.\cite{Barker76,Hansen86}
This is an approximate relation
between the pair and the direct correlation functions in addition to the exact Orstein--Zernike integral equation.
Unlike exact theories, these approximate closures introduce well known inconsistencies among different routes to
the equation of state. Given the knowledge of the pair correlation function, there are clearly  many
possible routes leading to the equation
of state, but the most frequently used are the energy, the virial (or pressure), and the compressibility routes.

The degree of inconsistency clearly depends upon the goodness of the approximate closure, so that some closures
might display weaker differences than others and might even, under some particular circumstances, give no difference
at all between two particular routes. The most notable example of this (although rarely mentioned in the literature) is
the virial-energy consistency within the hypernetted-chain (HNC) approximation.\cite{Morita60}
Other more recent examples
include the energy and virial routes in the hard-sphere limit of the square-shoulder potential (for any approximation),\cite{Santos05,Santos06}
and again the energy-virial consistency for soft-potentials within the mean-spherical approximation (MSA).\cite{Santos07}

The aim of this Communication is to add one more case to this relatively short list by showing that energy and virial
routes are completely equivalent within the linearized Debye--H\"uckel  (LDH) approximation for \textit{any
potential} in \textit{any} dimensionality.
Our interest in this problem has been triggered by recent investigations on bounded potentials,\cite{Malijevsky07,Santos08,Fantoni09}
where this consistency is of particular importance, as we shall discuss.

Consider an arbitrary potential $\phi(\mathbf{r})$ for a homogeneous fluid of $N$ particles in $d$ dimensions.
Newton's third law of motion implies that $\phi(\mathbf{r})=\phi(-\mathbf{r})$, but  the potential need not be spherically symmetric.
The virial equation is associated with the compressibility factor $Z(\rho,\beta)$ as\cite{Hansen86}
\begin{eqnarray}
\label{eq1}
Z \equiv \frac{\beta P}{\rho} &=& 1 + \frac{\rho}{2d}
\int d \mathbf{r}~ y\left(\mathbf{r}\right) \mathbf{r} \cdot \nabla f\left(\mathbf{r}\right),
\end{eqnarray}
where $\beta=1/k_B T$ is the inverse temperature, $\rho$ and $P$ are  the density and the pressure, respectively,
and $f(\mathbf{r};\beta)=e^{-\beta \phi(\mathbf{r})}-1$ is the Mayer function. In Eq.\ (\ref{eq1}) we have also
introduced the cavity function
$y(\mathbf{r};\rho,\beta)$, which is related to the pair correlation function $g(\mathbf{r};\rho,\beta)$ by the
relation
$y(\mathbf{r})=e^{\beta \phi(\mathbf{r})} g(\mathbf{r})$. The energy equation is associated with the energy $U$ per
particle $u(\rho,\beta)$,
\begin{eqnarray}
\label{eq2}
u \equiv \frac{U}{N} &=& \frac{d}{2\beta} -\frac{\rho}{2}  \int d \mathbf{r}~ y\left(\mathbf{r}\right)
\frac{\partial f\left(\mathbf{r}\right)}{\partial \beta}.
\end{eqnarray}
A standard thermodynamic identity\cite{note1} provides the consistency condition between the pressure and energy routes:
\begin{eqnarray}
\label{eq3}
\rho \left(\frac{\partial u}{\partial \rho} \right)_{\beta} &=& \left(\frac{\partial Z}{\partial \beta}\right)_{\rho}.
\end{eqnarray}
The explicit notation of the variable being kept fixed and the parameter dependence will be dropped henceforth for
notational simplicity.
It proves convenient to further introduce the quantity $w(\mathbf{r};\rho,\beta)=y(\mathbf{r};\rho,\beta)-1$, which is related to the potential of mean force $\psi(\mathbf{r};\rho,\beta)=-\beta^{-1}\ln g(\mathbf{r};\rho,\beta)$ by $-\beta[\psi(\mathbf{r})-\phi(\mathbf{r})]=\ln[1+w(\mathbf{r})]$. We anticipate that $w(\mathbf{r})$, introduced as a definition
at this stage, will assume a particular physical meaning within the LDH approximation later on.
Then, identity (\ref{eq3}) translates into the following condition:
\begin{eqnarray}
\label{eq4}
-\frac{\partial}{\partial \rho} \left[ \rho\int d \mathbf{r}~ w\left(\mathbf{r}\right) \frac{\partial
f\left(\mathbf{r}\right)}{\partial \beta} \right]
&=& \frac{1}{d} \frac{\partial}{\partial \beta} \left[\int d \mathbf{r}~ w\left(\mathbf{r}\right) \mathbf{r}
\cdot \nabla
f\left(\mathbf{r}\right) \right].
\end{eqnarray}
Introducing the Fourier transforms $\widetilde{w}(\mathbf{k};\rho,\beta)$ and $\widetilde{f}(\mathbf{k};\beta)$ of
$w(\mathbf{r};\rho,\beta)$
and $f(\mathbf{r};\beta)$, respectively, Eq.\ (\ref{eq4}) becomes, after standard manipulations,
\begin{eqnarray}
\label{eq5}
\frac{\partial}{\partial \rho} \left[ \rho \int \frac{d \mathbf{k}}{\left(2\pi\right)^d} ~\widetilde{w}
\left(\mathbf{k}\right) \frac{\partial \widetilde{f}\left(\mathbf{k}\right)}
{\partial \beta} \right] &=& \frac{\partial}{\partial \beta} \left [\frac{1}{d} \int \frac{d \mathbf{k}}
{\left(2\pi\right)^d}~
\widetilde{w}\left(\mathbf{k}\right) \nabla_{\mathbf{k}} \cdot \left( \mathbf{k} \widetilde{f}\left(\mathbf{k}\right) \right)
\right].
\end{eqnarray}
Here we have used $\widetilde{w}
\left(-\mathbf{k}\right)=\widetilde{w}
\left(\mathbf{k}\right)$ from the symmetry relation $\phi(-\mathbf{r})=\phi(\mathbf{r})$. Equation \eqref{eq5} can be recast into a more convenient form by taking into account the mathematical identity
\begin{eqnarray}
\label{eq6}
\frac{\partial}{\partial \beta} \left[ \widetilde{w}\left(\mathbf{k}\right) \nabla_{\mathbf{k}} \cdot \left(\mathbf{k}
\widetilde{f}\left(\mathbf{k}\right)\right) \right]&=&
d \frac{\partial \widetilde{w}\left(\mathbf{k}\right)}{\partial \beta} \widetilde{f}\left(\mathbf{k}\right) + \nabla_{\mathbf{k}} \cdot
\left[\mathbf{k}
\widetilde{w}\left(\mathbf{k}\right) \frac{\partial \widetilde{f} \left(\mathbf{k}\right)}{\partial \beta} \right]\nonumber\\
&&+
\mathbf{k}\cdot \left[\frac{\partial \widetilde{w}\left(\mathbf{k}\right)}{\partial \beta} \nabla_\mathbf{k} \widetilde{f}\left(\mathbf{k}\right)-
\frac{\partial \widetilde{f}\left(\mathbf{k}\right)}{\partial \beta} \nabla_\mathbf{k}\widetilde{w}\left(\mathbf{k}\right) \right].
\end{eqnarray}
Upon integration over $\mathbf{k}$ in Eq.\ (\ref{eq5}) the second surface term of the right-hand side of
Eq.\ (\ref{eq6}) can
be dropped and hence we find
\begin{eqnarray}
\label{eq7}
\frac{\partial}{\partial \rho} \left[ \rho \int \frac{d \mathbf{k}}{\left(2\pi\right)^d} ~\widetilde{w}
\left(\mathbf{k}\right) \frac{\partial \widetilde{f}\left(\mathbf{k}\right)}
{\partial \beta} \right] &=& \int \frac{d \mathbf{k}}{\left(2\pi\right)^d} ~\frac{\partial \widetilde{w}
\left(\mathbf{k}\right)}{\partial \beta}
\widetilde{f} \left(\mathbf{k}\right) \nonumber\\
&&+ \frac{1}{d} \int \frac{d \mathbf{k}}{\left(2\pi\right)^d}~
\mathbf{k}\cdot \left[\frac{\partial \widetilde{w}\left(\mathbf{k}\right)}{\partial \beta} \nabla_\mathbf{k} \widetilde{f}\left(\mathbf{k}\right)-
\frac{\partial \widetilde{f}\left(\mathbf{k}\right)}{\partial \beta} \nabla_\mathbf{k}\widetilde{w}\left(\mathbf{k}\right) \right].
\end{eqnarray}
We remark that no approximations have been carried out so far, and that Eq.\ (\ref{eq7}) is completely equivalent
to the consistency
condition (\ref{eq3}). Therefore, \textit{any} $\widetilde{w}(\mathbf{k})$ satisfying Eq.\ (\ref{eq7}) gives thermodynamically
consistent results
via the energy and virial routes.

We now show that this is in fact the case for the LDH theory which is defined by $w(\mathbf{r})=y(\mathbf{r})-1$ where
$\widetilde{w}(\mathbf{k})$ satisfies the scaling relation\cite{note2}
\begin{eqnarray}
\label{eq8}
\rho \widetilde{w}\left(\mathbf{k}\right) &=& F\left(\rho \widetilde{f}\left(\mathbf{k}\right)\right),
\end{eqnarray}
with $F\left(z\right) = {z^2}/({1-z})$.
This immediately provides the following expressions
\begin{eqnarray}
\label{eq10}
\frac{\partial}{\partial \rho} \left[\rho \widetilde{w}\left(\mathbf{k}\right)\right]&=& F'\left(\rho \widetilde{f}\left(\mathbf{k}\right)\right)
\widetilde{f}\left(\mathbf{k}\right), \\
\label{eq11}
\frac{\partial \widetilde{w}\left(\mathbf{k}\right)}{\partial \beta} &=&F'\left(\rho \widetilde{f}\left(\mathbf{k}\right)\right)
\frac{\partial \widetilde{f}\left(\mathbf{k}\right)}{\partial \beta}, \\
\label{eq12}
\nabla_\mathbf{k} \widetilde{w}\left(\mathbf{k}\right)&=&F'\left(\rho \widetilde{f}\left(\mathbf{k}\right)\right)
\nabla_\mathbf{k} \widetilde{f}\left(\mathbf{k}\right).
\end{eqnarray}
Equations (\ref{eq11}) and (\ref{eq12}) readily yield
\begin{eqnarray}
\label{eq13}
\frac{\partial \widetilde{w}\left(\mathbf{k}\right)}{\partial \beta} \nabla_\mathbf{k} \widetilde{f}\left(\mathbf{k}\right)&=&
\frac{\partial \widetilde{f}\left(\mathbf{k}\right)}{\partial \beta} \nabla_\mathbf{k}\widetilde{w}\left(\mathbf{k}\right),
\end{eqnarray}
so that the second integral on the right-hand side of (\ref{eq7}) vanishes identically. In addition, Eqs.\
(\ref{eq10}) and \eqref{eq11} show that the remaining terms in Eq.\ (\ref{eq7}) are  identical. This closes the proof.

It is useful to put the present result into some perspectives. The LDH theory can be derived from diagrammatic
methods\cite{Hansen86} by summing all simple chain diagrams to all orders in density $\rho$. A mathematical device to
do this is to formally multiply the Mayer function $f(\mathbf{r})$ by a bookkeeping parameter $\mu$, and then let $\mu \to 0$,
so that the leading diagrams to be retained at each order are the simple  chain diagrams, which then give the
dominant contribution to the pair correlation function within this approximation. This procedure is physically
justified only for \textit{bounded} potentials where $|f(\mathbf{r})|$ can be made arbitrarily small
by increasing the temperature,
and hence the virial-energy consistency is also representative of the exact behavior of the equation of state, unlike
the case of unbounded potentials where this is not the case and consistency does not automatically ensure exact results.\cite{note4}

Representative examples of bounded potentials, recently discussed in the literature, include Gaussian potentials,\cite{Mladek06} penetrable spheres (PS),\cite{Malijevsky07} and penetrable square-well (PSW).\cite{Santos08,Fantoni09}
These potentials are currently of the greatest interest both from a practical point of view, as they mimic ultrasoft systems
such as suitable mixtures of colloids and polymers,\cite{Likos01} and theoretically, as they are compatible with
a phase transition even in one-dimensional systems (see for instance discussion in Ref.\ \onlinecite{Fantoni09}).

\begin{figure}[h]
\includegraphics[width=10cm]{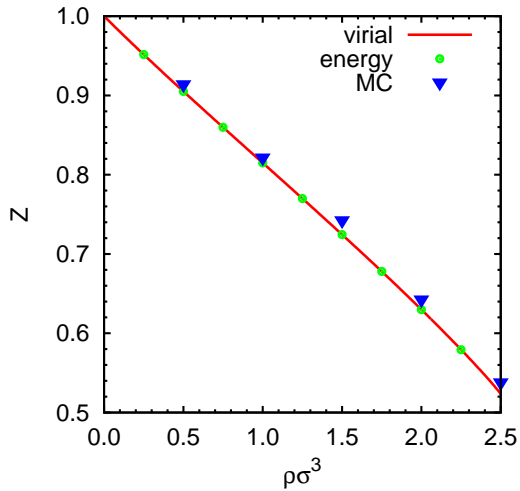}
\caption{Equation of state of the three-dimensional PSW model, Eq.\ \protect\eqref{eq14}, as obtained from the virial route, Eq.\ (\protect\ref{eq1}), (solid line) and from
the energy route, Eq.\ (\protect\ref{eq2}), (solid circles) for reduced temperature $k_{B} T/\epsilon_a=8$, well width
$\Delta/\sigma=0.5$, and energy ratio $\epsilon_r/\epsilon_a=2$. Also shown are the results obtained from MC simulations (triangles).
\label{fig:fig1}}
\end{figure}

We have explicitly numerically checked the virial-energy equation within the PSW model defined by the potential\cite{Santos08,Fantoni09}
\begin{eqnarray}
\label{eq14}
\phi\left(r\right)=\left\{
\begin{array}{ll}
\epsilon_r, & r<\sigma,\\
-\epsilon_a, & \sigma<r<\sigma+\Delta,\\
0 ,          & r>\sigma+\Delta,
\end{array}\right.
\end{eqnarray}
where $\sigma$ is the particle diameter, $\Delta$ is the width of the well, and $\epsilon_r$ and $\epsilon_a$ are two positive
constants accounting for the repulsive and attractive parts of the potential, respectively. Two particles then attract each
other through
a square-well potential of depth $-\epsilon_a<0$ and width $\Delta$, but can also interpenetrate each other with an energy cost
$\epsilon_r>0$. Figure \ref{fig:fig1} depicts both the virial and the energy equation of state for the PSW model at a representative state point,
as obtained from a numerical solution  within the LDH approximation.
As expected, we find complete numerical consistency, in agreement with the analytical proof. The compressibility equation of state (not shown) lies slightly below the energy-virial curve.
Figure \ref{fig:fig1} also includes Monte Carlo (MC) data obtained for the same system and state.\cite{MC} We observe that at this relatively high temperature the LDH solution  provides an accurate equation of state, in agreement with the previous discussion on bounded potentials.

We close this Communication with a few remarks. The analytical proof presented here is patterned after a similar proof on the
virial-energy consistency within the MSA for a general class of soft potentials which include bounded
interactions treated here.\cite{Santos07} Given the close relationship between MSA and LDH for soft potentials,\cite{Hansen86}
the result presented here and in Ref.\ \onlinecite{Santos07} retrospectively can be cast within a unified framework associated
with the existence of a scaling form in Fourier space akin to Eq.\ (\ref{eq8}).

Of different nature appears to be the virial-energy consistency within the HNC closure. This is a direct consequence of
the existence of an explicit expression  for the free energy, pressure, and chemical potential as a result of a \textit{single}
approximation, thus increasing internal consistency.\cite{Morita60,Schlijper93,Olivares76}

On the other hand, the HNC theory can be alternatively viewed as an approximation to the exact diagrammatic expansion of the
pair correlation function which retains the complete class of particular diagrams (chains, both simple and netted, and bundles)
and the virial-energy consistency can be also regarded as a direct consequence of this.\cite{Barker76,Balescu75} As the full
expansion including all diagrams is of course  consistent, an additional further consequence is that  the class of
diagrams not included within the HNC approximation (the so-called elementary diagrams related to the bridge function)
must also be consistent from the virial-energy point of view. Our result builds upon this argument by adding the
additional piece of information that the full inclusion of simple chain diagrams  only also leads to virial-energy
consistency. A profound consequence of our result is therefore  that the virial-energy consistency is
deeply tied to the retention of all diagrams within a given class.\cite{note3}
\begin{acknowledgments}
The research of A.S. was supported by the Ministerio de Educaci\'on y Ciencia (Spain) through Grant No.\
FIS2007-60977 (partially financed by FEDER funds) and by the Junta de Extremadura through Grant No.\ GRU09038.
The work of R.F. and A.G. was supported by the Italian MIUR through a grant PRIN-COFIN 2007B57EAB(2008/2009).
The authors are grateful to D. Henderson and F. Lado for a critical reading of the manuscript.
\end{acknowledgments}


\begin{thebibliography}{99}
\bibitem{note1}
This can be quickly established by starting from the definition of the Helmholtz free energy $F=U-TS$
($S$ being the entropy), so that $P=-(\partial F/\partial V)_{\beta,N}$,
$Z=\rho [\partial(\beta F/N)/\partial \rho]_{\beta}$, and $U=[\partial (\beta F)/\partial \beta]_{N,V}$.
Equation (\ref{eq3}) then follows immediately.
\bibitem{note4}
Note that the LDH theory was originally devised for Coulomb potential where it is justified for distances
larger than the Debye length (and sufficiently high temperatures). For bounded potentials
(and again sufficiently high temperatures), the region of reliability is extended to all distances.
\bibitem{MC}
The MC data were generated  from standard $NVT$ simulations of 108 particles with 8 million MC steps (one particle move per step).
The pressure was determined from the virial equation through analysis of the pair correlation function at the singularities of
the potential.
\bibitem{note3}
In this respect, note that the Percus--Yevick theory, which retains all linear chains but only a subset of netted chains
in the diagrammatic expansion of $y(\mathbf{r})$, yields in general inconsistent energy and virial equations of state.
\end{thebibliography}

\end{document}